\documentclass[aps,prl,twocolumn,showpacs,floatfix]{revtex4}

\usepackage{graphicx}

\newcommand{\met}       {\mbox{$\not\!\!E_T$}}
\newcommand{\rar}       {\rightarrow}
\newcommand{\rargap}    {\mbox{ $\rightarrow$ }}
\newcommand{\ttbar}     {\mbox{$t\bar{t}$}}
\newcommand{\ppbar}     {\mbox{$p\bar{p}$}}

\begin{document}

\title{Evidence for production of single top quarks and
first direct measurement of \boldmath$|V_{tb}|$}

% LIST_OF_AUTHORS_R2.TEX                11/21/06            
%
\author{                                                                      
%% names begin here                                                           
V.M.~Abazov,$^{35}$                                                           
B.~Abbott,$^{75}$                                                             
M.~Abolins,$^{65}$                                                            
B.S.~Acharya,$^{28}$                                                          
M.~Adams,$^{51}$                                                              
T.~Adams,$^{49}$                                                              
E.~Aguilo,$^{5}$                                                              
S.H.~Ahn,$^{30}$                                                              
M.~Ahsan,$^{59}$                                                              
G.D.~Alexeev,$^{35}$
% ...
\\                                                                            
\vspace{0.1in}                                                                 
\centerline{(D\O\ Collaboration)}                                             
}                                                                             
\affiliation{                                                                 
\centerline{$^{1}$Universidad de Buenos Aires, Buenos Aires, Argentina}       
\centerline{$^{2}$LAFEX, Centro Brasileiro de Pesquisas F{\'\i}sicas,         
                  Rio de Janeiro, Brazil}                                     
\centerline{$^{3}$Universidade do Estado do Rio de Janeiro,                   
                  Rio de Janeiro, Brazil}                                     
\centerline{$^{4}$Instituto de F\'{\i}sica Te\'orica, Universidade            
                  Estadual Paulista, S\~ao Paulo, Brazil}                     
\centerline{$^{5}$University of Alberta, Edmonton, Alberta, Canada,           
                  Simon Fraser University, Burnaby, British Columbia, Canada,}
\centerline{York University, Toronto, Ontario, Canada, and                    
                  McGill University, Montreal, Quebec, Canada}                
\centerline{$^{6}$University of Science and Technology of China, Hefei,       
                  People's Republic of China}                                 
\centerline{$^{7}$Universidad de los Andes, Bogot\'{a}, Colombia}             
\centerline{$^{8}$Center for Particle Physics, Charles University,            
                  Prague, Czech Republic}                                     
\centerline{$^{9}$Czech Technical University, Prague, Czech Republic}         
\centerline{$^{10}$Center for Particle Physics, Institute of Physics,         
                   Academy of Sciences of the Czech Republic,                 
                   Prague, Czech Republic}                                    
\centerline{$^{11}$Universidad San Francisco de Quito, Quito, Ecuador}        
\centerline{$^{12}$Laboratoire de Physique Corpusculaire, IN2P3-CNRS,         
                   Universit\'e Blaise Pascal, Clermont-Ferrand, France}      
\centerline{$^{13}$Laboratoire de Physique Subatomique et de Cosmologie,      
                   IN2P3-CNRS, Universite de Grenoble 1, Grenoble, France}    
\centerline{$^{14}$CPPM, IN2P3-CNRS, Universit\'e de la M\'editerran\'ee,     
                   Marseille, France}                                         
\centerline{$^{15}$Laboratoire de l'Acc\'el\'erateur Lin\'eaire,              
                   IN2P3-CNRS et Universit\'e Paris-Sud, Orsay, France}       
\centerline{$^{16}$LPNHE, IN2P3-CNRS, Universit\'es Paris VI and VII,         
                   Paris, France}                                             
\centerline{$^{17}$DAPNIA/Service de Physique des Particules, CEA, Saclay,    
                   France}                                                    
\centerline{$^{18}$IPHC, IN2P3-CNRS, Universit\'e Louis Pasteur, Strasbourg,  
                   France, and Universit\'e de Haute Alsace,                  
                   Mulhouse, France}                                          
\centerline{$^{19}$Institut de Physique Nucl\'eaire de Lyon, IN2P3-CNRS,      
                   Universit\'e Claude Bernard, Villeurbanne, France}         
\centerline{$^{20}$III. Physikalisches Institut A, RWTH Aachen,               
                   Aachen, Germany}                                           
\centerline{$^{21}$Physikalisches Institut, Universit{\"a}t Bonn,             
                   Bonn, Germany}                                             
\centerline{$^{22}$Physikalisches Institut, Universit{\"a}t Freiburg,         
                   Freiburg, Germany}                                         
\centerline{$^{23}$Institut f{\"u}r Physik, Universit{\"a}t Mainz,            
                   Mainz, Germany}                                            
\centerline{$^{24}$Ludwig-Maximilians-Universit{\"a}t M{\"u}nchen,            
                   M{\"u}nchen, Germany}                                      
\centerline{$^{25}$Fachbereich Physik, University of Wuppertal,               
                   Wuppertal, Germany}                                        
\centerline{$^{26}$Panjab University, Chandigarh, India}                      
\centerline{$^{27}$Delhi University, Delhi, India}                            
\centerline{$^{28}$Tata Institute of Fundamental Research, Mumbai, India}     
\centerline{$^{29}$University College Dublin, Dublin, Ireland}                
\centerline{$^{30}$Korea Detector Laboratory, Korea University,               
                   Seoul, Korea}                                              
\centerline{$^{31}$SungKyunKwan University, Suwon, Korea}                     
\centerline{$^{32}$CINVESTAV, Mexico City, Mexico}                            
\centerline{$^{33}$FOM-Institute NIKHEF and University of                     
                   Amsterdam/NIKHEF, Amsterdam, The Netherlands}              
\centerline{$^{34}$Radboud University Nijmegen/NIKHEF, Nijmegen, The          
                  Netherlands}                                                
\centerline{$^{35}$Joint Institute for Nuclear Research, Dubna, Russia}       
\centerline{$^{36}$Institute for Theoretical and Experimental Physics,        
                   Moscow, Russia}                                            
\centerline{$^{37}$Moscow State University, Moscow, Russia}                   
\centerline{$^{38}$Institute for High Energy Physics, Protvino, Russia}       
\centerline{$^{39}$Petersburg Nuclear Physics Institute,                      
                   St. Petersburg, Russia}                                    
\centerline{$^{40}$Lund University, Lund, Sweden, Royal Institute of          
                   Technology and Stockholm University, Stockholm,            
                   Sweden, and}                                               
\centerline{Uppsala University, Uppsala, Sweden}                              
\centerline{$^{41}$Physik Institut der Universit{\"a}t Z{\"u}rich,            
                   Z{\"u}rich, Switzerland}                                   
\centerline{$^{42}$Lancaster University, Lancaster, United Kingdom}           
\centerline{$^{43}$Imperial College, London, United Kingdom}                  
\centerline{$^{44}$University of Manchester, Manchester, United Kingdom}      
\centerline{$^{45}$University of Arizona, Tucson, Arizona 85721, USA}         
\centerline{$^{46}$Lawrence Berkeley National Laboratory and University of    
                   California, Berkeley, California 94720, USA}               
\centerline{$^{47}$California State University, Fresno, California 93740, USA}
\centerline{$^{48}$University of California, Riverside, California 92521, USA}
\centerline{$^{49}$Florida State University, Tallahassee, Florida 32306, USA} 
\centerline{$^{50}$Fermi National Accelerator Laboratory,                     
            Batavia, Illinois 60510, USA}                                     
\centerline{$^{51}$University of Illinois at Chicago,                         
            Chicago, Illinois 60607, USA}                                     
\centerline{$^{52}$Northern Illinois University, DeKalb, Illinois 60115, USA} 
\centerline{$^{53}$Northwestern University, Evanston, Illinois 60208, USA}    
\centerline{$^{54}$Indiana University, Bloomington, Indiana 47405, USA}       
\centerline{$^{55}$University of Notre Dame, Notre Dame, Indiana 46556, USA}  
\centerline{$^{56}$Purdue University Calumet, Hammond, Indiana 46323, USA}    
\centerline{$^{57}$Iowa State University, Ames, Iowa 50011, USA}              
\centerline{$^{58}$University of Kansas, Lawrence, Kansas 66045, USA}         
\centerline{$^{59}$Kansas State University, Manhattan, Kansas 66506, USA}     
\centerline{$^{60}$Louisiana Tech University, Ruston, Louisiana 71272, USA}   
\centerline{$^{61}$University of Maryland, College Park, Maryland 20742, USA} 
\centerline{$^{62}$Boston University, Boston, Massachusetts 02215, USA}       
\centerline{$^{63}$Northeastern University, Boston, Massachusetts 02115, USA} 
\centerline{$^{64}$University of Michigan, Ann Arbor, Michigan 48109, USA}    
\centerline{$^{65}$Michigan State University,                                 
            East Lansing, Michigan 48824, USA}                                
\centerline{$^{66}$University of Mississippi,                                 
            University, Mississippi 38677, USA}                               
\centerline{$^{67}$University of Nebraska, Lincoln, Nebraska 68588, USA}      
\centerline{$^{68}$Princeton University, Princeton, New Jersey 08544, USA}    
\centerline{$^{69}$State University of New York, Buffalo, New York 14260, USA}
\centerline{$^{70}$Columbia University, New York, New York 10027, USA}        
\centerline{$^{71}$University of Rochester, Rochester, New York 14627, USA}   
\centerline{$^{72}$State University of New York,                              
            Stony Brook, New York 11794, USA}                                 
\centerline{$^{73}$Brookhaven National Laboratory, Upton, New York 11973, USA}
\centerline{$^{74}$Langston University, Langston, Oklahoma 73050, USA}        
\centerline{$^{75}$University of Oklahoma, Norman, Oklahoma 73019, USA}       
\centerline{$^{76}$Oklahoma State University, Stillwater, Oklahoma 74078, USA}
\centerline{$^{77}$Brown University, Providence, Rhode Island 02912, USA}     
\centerline{$^{78}$University of Texas, Arlington, Texas 76019, USA}          
\centerline{$^{79}$Southern Methodist University, Dallas, Texas 75275, USA}   
\centerline{$^{80}$Rice University, Houston, Texas 77005, USA}                
\centerline{$^{81}$University of Virginia, Charlottesville,                   
            Virginia 22901, USA}                                              
\centerline{$^{82}$University of Washington, Seattle, Washington 98195, USA}
\vspace{-0.2in}                                                     
}                                                                             
%end                                                                          

\date{April 9, 2007}

\begin{abstract}
The D0 Collaboration presents first evidence for the production of
single top quarks at the Fermilab Tevatron {\ppbar} collider. Using a
0.9~fb$^{-1}$ dataset, we apply a multivariate analysis to separate
signal from background and measure
$\sigma({\ppbar}{\rargap}tb+X,~tqb+X) = 4.9 \pm 1.4$~pb. The
probability to measure a cross section at this value or higher in the
absence of signal is $0.035\%$, corresponding to a 3.4~standard
deviation significance. We use the cross section measurement to
directly determine the CKM matrix element that describes the $Wtb$
coupling and find $0.68 < |V_{tb}| \le 1$ at $95\%$~C.L. within the
standard model.
\end{abstract}

\pacs{14.65.Ha; 12.15.Ji; 13.85.Qk \vspace{-0.25in}}

\maketitle 

%---------------------------------------------------------------------
%---------------------------------------------------------------------

\begin{figure}[!h!tbp]
\includegraphics[width=3.3in]{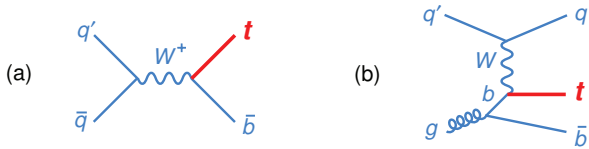}
\vspace{-0.12in}
\caption[feynman]{Representative Feynman diagrams for (a) s-channel
single top quark production and (b) t-channel production.}
\label{feynman}
\end{figure}

Top quarks were first observed in strong {\ttbar} pair production at
the Tevatron collider in 1995~\cite{top-obs-1995}. In the standard
model (SM), $\sigma({\ppbar}{\rargap}{\ttbar}+X) = 6.8 ^{+0.6}
_{-0.5}$~pb~\cite{ttbar-xsec-1} at $\sqrt{s} = 1.96$~TeV for a top
quark mass of 175~GeV. Top quarks are also expected to be produced
singly via the electroweak
processes~\cite{singletop-theory,singletop-yuan} illustrated in
Fig.~\ref{feynman}. For brevity, we use the notation ``$tb$'' to
represent the sum of $t\bar{b}$ and $\bar{t}b$, and ``$tqb$'' for the
sum of $tq\bar{b}$ and $\bar{t}\bar{q}b$. The next-to-leading order
(NLO) prediction for the s-channel single top quark cross section is
$\sigma({\ppbar}{\rargap}tb+X) = 0.88\pm0.11$~pb, and for the
t-channel process, the prediction is $\sigma({\ppbar} {\rargap}tqb+X)
= 1.98\pm0.25$~pb~\cite{singletop-xsec-sullivan,singletop-xsec-kidonakis}.

\vspace{-0.1in}
\vspace{-0.1in}

Single top quark events can be used to study the $Wtb$
coupling~\cite{singletop-wtb-heinson}, and to measure the magnitude of
the element $|V_{tb}|$ of the quark mixing matrix, (the
Cabibbo-Kobayashi-Maskawa (CKM) matrix~\cite{ckm-matrix}), without
assuming only three generations of
quarks~\cite{singletop-vtb-jikia}. The quark mixing matrix must be
unitary, which for three families implies $|V_{tb}| \simeq
1$~\cite{pdb-vtb}. A smaller measured value would indicate the
presence of a fourth quark family to make up the difference. Single
top quark production can also be used to measure the top quark partial
decay width $\Gamma(t{\rar}Wb)$~\cite{singletop-width-yuan} and hence
the top quark lifetime.

The D0 collaboration has previously published limits~\cite{limits-d0}
on single top quark production. The best $95\%$~C.L. upper limits are
$\sigma({\ppbar}{\rargap}tb+X)<6.4$~pb and
$\sigma({\ppbar}{\rargap}tqb+X)<5.0$~pb. The CDF collaboration has
also published limits on the cross sections~\cite{limits-cdf}.

This Letter describes a search for single top quark production using
0.9~fb$^{-1}$ of data produced at a center-of-mass energy of 1.96~TeV.
The data were collected from 2002 to 2005 using the D0
detector~\cite{d0-detector} with triggers that required a jet and an
electron or a muon. The search focuses on the final state consisting
of one high transverse momentum ($p_T$) isolated lepton and missing
transverse energy ({\met}), together with a $b$-quark jet from the
decay of the top quark ($t{\rar}Wb{\rar}\ell \nu b$). There is an
additional $b$~quark in s-channel production, and an additional light
quark and $b$~quark in t-channel production. The second $b$~quark in
the t-channel is rarely reconstructed since it is produced in the
forward direction with low transverse momentum. The main backgrounds
are: $W$~bosons produced in association with jets; top quark pairs
decaying into the lepton+jets and dilepton final states, when a jet or
a lepton is not reconstructed; and multijet production, where a jet is
misreconstructed as an electron, or a heavy-flavor quark decays to a
muon that passes the isolation criteria.

We model the signal using the {\sc singletop} NLO Monte Carlo (MC)
event generator~\cite{singletop-mcgen}. The event kinematics for both
s-channel and t-channel reproduce distributions found in NLO
calculations~\cite{singletop-xsec-sullivan}. The decays of the top
quark and resulting $W$~boson, with finite widths, are modeled in the
{\sc singletop} generator to preserve particle spin information. {\sc
Pythia}~\cite{pythia} is used to model the hadronization of the
generated partons. For the $tb$ search, we assume SM $tqb$ as part of
the background, and vice versa. For the $tb$+$tqb$ search, we assume
the SM ratio between the $tb$ and $tqb$ cross sections.

We simulate the {\ttbar} and $W$+jets backgrounds using the {\sc
alpgen} leading-order MC event generator~\cite{alpgen} and {\sc
pythia} to model the hadronization. A parton-jet matching
algorithm~\cite{jet-matching} is used to ensure there is no
double-counting of the final states. The {\ttbar} background is
normalized to the integrated luminosity times the predicted {\ttbar}
cross section~\cite{ttbar-xsec-1}. The multijet background is modeled
using data that contain nonisolated leptons but which otherwise
resemble the lepton+jets dataset. The $W$+jets background, combined
with the multijet background, is normalized to the lepton+jets dataset
separately for each analysis channel (defined by lepton flavor and jet
multiplicity) before $b$-jet tagging (described later). In the
$W$+jets background simulation, we scale the $Wb\bar{b}$ and
$Wc\bar{c}$ components by a factor of $1.50\pm0.45$ to better
represent higher-order effects~\cite{wbb-wbj-nlo}. This factor is
determined by scaling the numbers of events in an admixture of light-
and heavy-flavor $W$+jets MC events to data that have no $b$~tags but
which otherwise pass all selection cuts. The uncertainty assigned to
this factor covers the expected dependence on kinematics and the
assumption that the factor is the same for $Wb\bar{b}$ and
$Wc\bar{c}$.

We pass the MC events through a {\sc geant}-based
simulation~\cite{geant} of the D0 detector. To correct differences
between the simulation and data, we apply weights to the simulated
events to model the effects of the triggers, lepton identification and
isolation requirements, and the energy scale of the jets. The
$b$-tagging algorithm~\cite{btagging-scanlon} is modeled by applying
weights that account for the probability for each jet to be tagged as
a function of jet flavor, $p_T$, and pseudorapidity $\eta$.

We choose events with two, three, or four jets, reconstructed using a
cone algorithm~\cite{cone-algorithm} with radius $\mathcal{R} =
\sqrt{(\Delta y)^2+(\Delta\phi)^2} = 0.5$ (where $y$ is rapidity and
$\phi$ is azimuthal angle) to cluster energy deposits in the
calorimeter. The leading jet has $p_T>25$~GeV and $|\eta|<2.5$, the
second leading jet has $p_T>20$~GeV and $|\eta|<3.4$, and subsequent
jets have $p_T>15$~GeV and $|\eta|<3.4$. Events are required to have
$15 < \met < 200$~GeV and exactly one isolated electron with
$p_T>15$~GeV and $|\eta|<1.1$ or one isolated muon with $p_T>18$~GeV
and $|\eta|<2.0$. Misreconstructed events are rejected by requiring
that the direction of the $\met$ is not aligned or anti-aligned in
azimuth with the lepton or a jet. To enhance the signal content of the
selection, one or two of the jets are required to be identified as
originating from long-lived $b$~hadrons by a neural network $b$-jet
tagging algorithm. The variables used to identify such jets rely on
the presence and characteristics of a secondary vertex and tracks with
high impact parameters inside the jet. For a $0.5\%$ light-jet $b$-tag
efficiency (the average mistag probability), we obtain a $50\%$
average tag rate in data for $b$~jets with $|\eta|<2.4$.

We select 1,398 $b$-tagged lepton+jets data events, which we expect to
contain $62\pm 13$ single top quark events. To increase the search
sensitivity, we divide these events into twelve independent analysis
channels based on the lepton flavor ($e$ or $\mu$), jet multiplicity
(2, 3, or 4), and number of identified $b$~jets (1 or 2). We do this
because the signal acceptance and signal-to-background ratio differ
significantly from channel to channel. Event yields are given in
Table~\ref{event-yields}, shown separated only by jet multiplicity for
simplicity. The acceptances for single top quark signal as percentages
of the total production cross sections are $(3.2\pm0.4)\%$ for $tb$
and $(2.1\pm0.3)\%$ for $tqb$.

The dominant contributions to the uncertainties on the backgrounds
come from: normalization of the {\ttbar} background (18\% of the
{\ttbar} component), which includes a term to account for the top
quark mass uncertainty; normalization of the $W$+jets and multijet
backgrounds to data (17--27\%), which includes the uncertainty on the
heavy-flavor fraction of the model; the jet energy scale corrections
(1--20\%); and the $b$-tagging probabilities (12--17\% for
double-tagged events). The uncertainty on the integrated luminosity is
6\%; all other sources contribute at the few percent level. The
uncertainties from the jet energy scale corrections and the
$b$-tagging probabilities affect both the shape and normalization of
the simulated distributions. Having selected the data samples, we
check that the background model reproduces the data in a multitude of
variables (e.g., transverse momenta, pseudorapidities, azimuthal
angles, masses) for each analysis channel and find agreement within
uncertainties.

\begin{table}[!h!tbp]
\vspace{-0.1in}
\caption[eventyields]{Numbers of expected and observed events in
0.9~fb$^{-1}$ for $e$ and $\mu$, 1 $b$~tag and 2 $b$~tag channels
combined. The total background uncertainties are smaller than the
component uncertainties added in quadrature because of anticorrelation
between the $W$+jets and multijet backgrounds resulting from the
background normalization procedure.}
\label{event-yields}
\begin{ruledtabular}
\begin{tabular}{l@{\extracolsep{\fill}}r@{\extracolsep{0pt}$\pm$}l@{}%
                 @{\extracolsep{\fill}}r@{\extracolsep{0pt}$\pm$}l@{}%
                 @{\extracolsep{\fill}}r@{\extracolsep{0pt}$\pm$}l@{}}
Source           & \multicolumn{2}{c}{2 jets}
                 & \multicolumn{2}{c}{3 jets}
                 & \multicolumn{2}{c}{4 jets} \\
\hline\hline	     
$tb$                      &  16  &   3  &   8  &  2  &   2  &  1  \\
$tqb$                     &  20  &   4  &  12  &  3  &   4  &  1  \\
\hline                                                              
${\ttbar}{\rar}\ell\ell$  &  39  &   9  &  32  &  7  &  11  &  3  \\
${\ttbar}{\rar}\ell$+jets &  20  &   5  & 103  & 25  & 143  & 33  \\
$Wb\bar{b}$               & 261  &  55  & 120  & 24  &  35  &  7  \\
$Wc\bar{c}$               & 151  &  31  &  85  & 17  &  23  &  5  \\
$Wjj$                     & 119  &  25  &  43  &  9  &  12  &  2  \\
Multijets                 &  95  &  19  &  77  & 15  &  29  &  6  \\
\hline                                                              
Total background          & 686  &  41  & 460  & 39  & 253  & 38  \\
Data             & \multicolumn{2}{c}{697}
                 & \multicolumn{2}{c}{455}
                 & \multicolumn{2}{c}{246}
\end{tabular}
\end{ruledtabular}
\vspace{-0.2in}
\end{table}

Since we expect single top quark events to constitute only a small
fraction of the selected event samples, and since the background
uncertainty is larger than the expected signal, a counting experiment
will not have sufficient sensitivity to verify their presence. We
proceed instead to calculate multivariate discriminants that separate
the signal from background and thus enhance the probability to observe
single top quarks. We use decision trees~\cite{decision-trees} to
create these discriminants. A decision tree is a machine-learning
technique that applies cuts iteratively to classify events. The
discrimination power is further improved by averaging over many
decision trees constructed using the adaptive boosting algorithm
AdaBoost~\cite{adaboost}. We refer to this average as a boosted
decision tree.

We identify 49 variables from an analysis of the signal and background
Feynman diagrams~\cite{singletop-vars-dudko-boos}, studies of single
top quark production at NLO~\cite{singletop-vars-cao}, and from other
analyses~\cite{singletop-yuan,singletop-vars-mahlon-parke}. The
variables may be classified into three categories: individual object
kinematics, global event kinematics, and variables based on angular
correlations. Those with the most discrimination power include the
invariant mass of all the jets in the event, the invariant mass of the
reconstructed $W$~boson and the highest-$p_T$ $b$-tagged jet, the
angle between the highest-$p_T$ $b$-tagged jet and the lepton in the
rest frame of the reconstructed top quark, and the lepton charge times
the pseudorapidity of the untagged jet. We find that reducing the
number of variables always reduces the sensitivity of the analysis.

\begin{figure}[!h!tbp]
\includegraphics[width=1.6in]{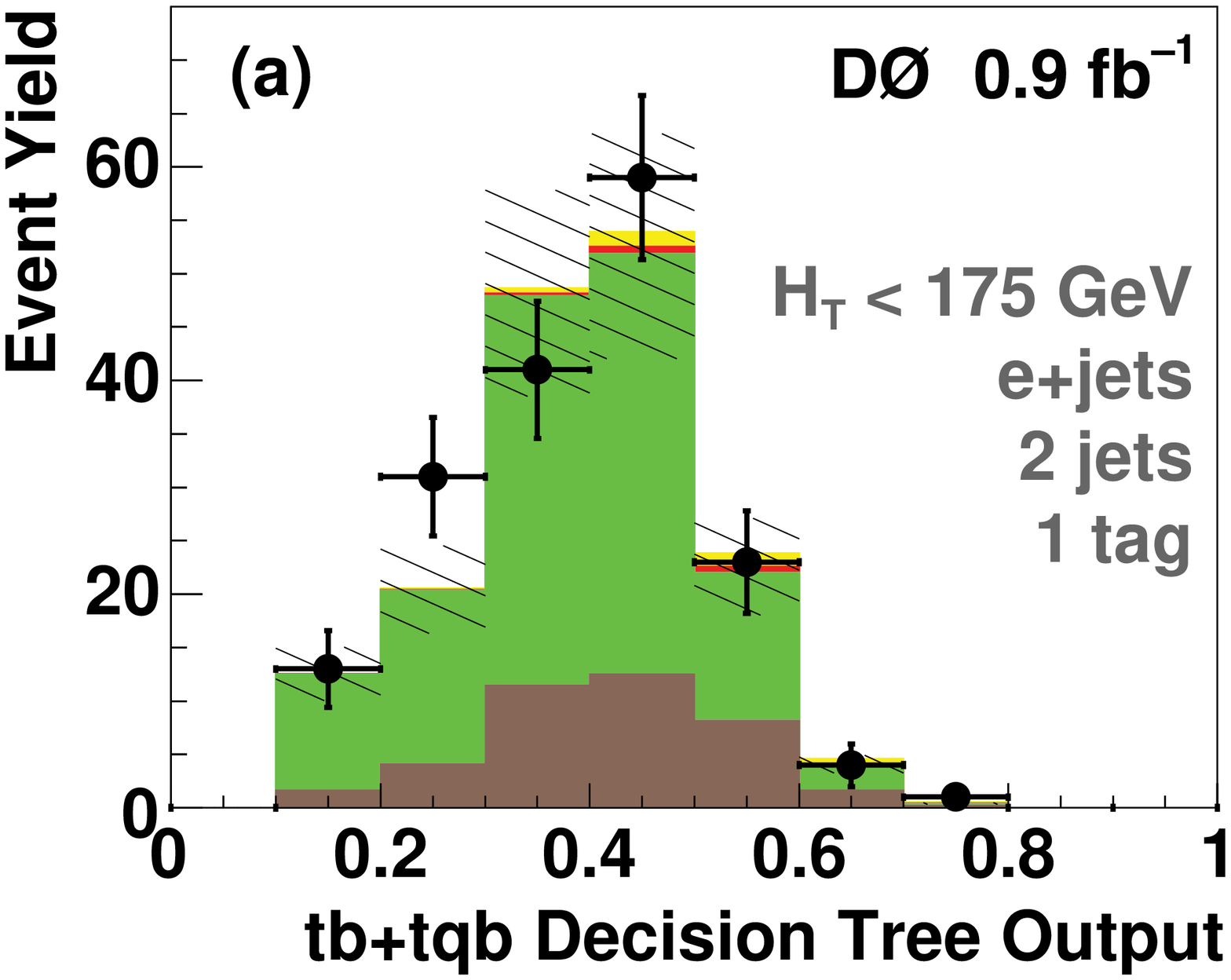}
\includegraphics[width=1.6in]{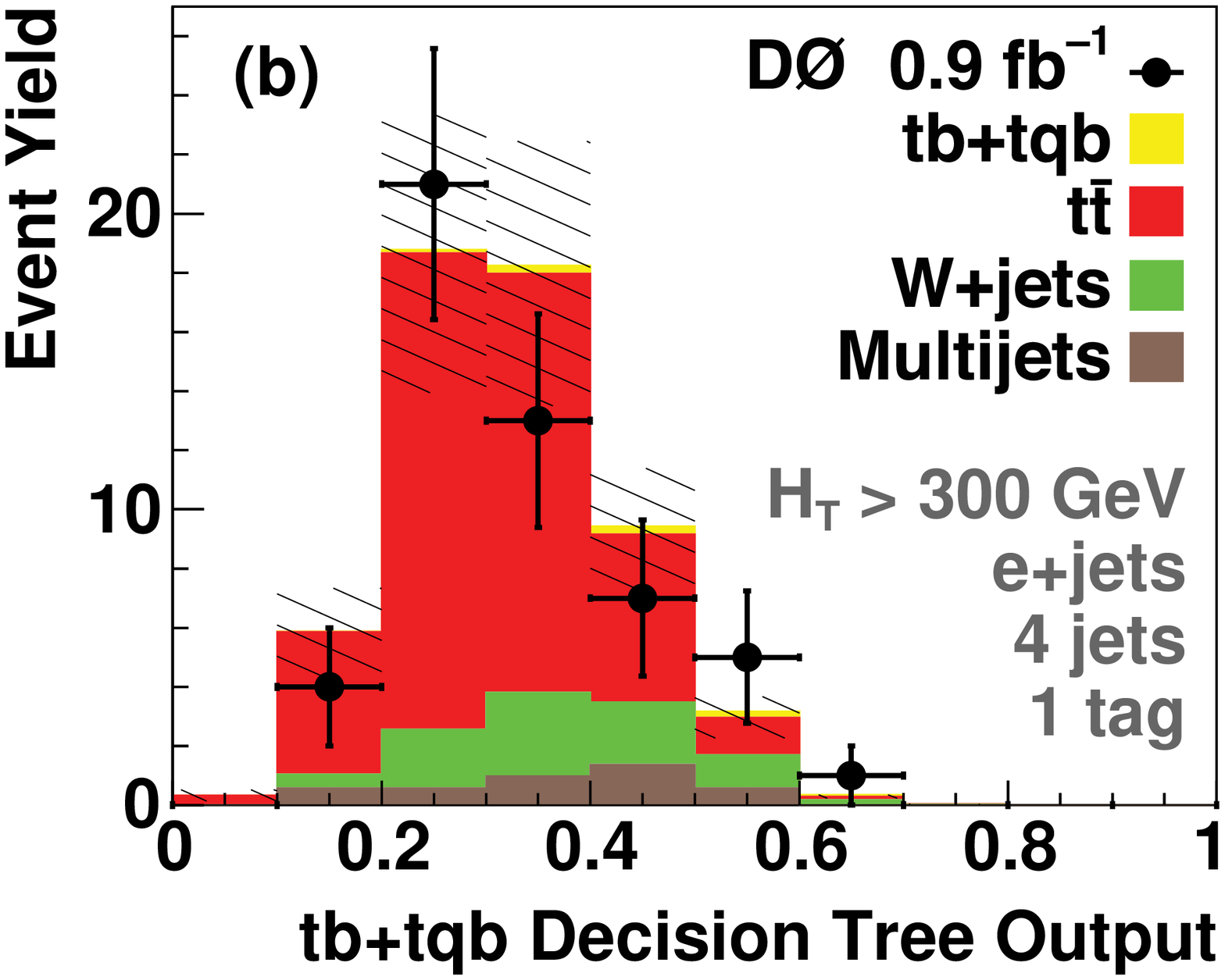}
\includegraphics[width=1.6in]{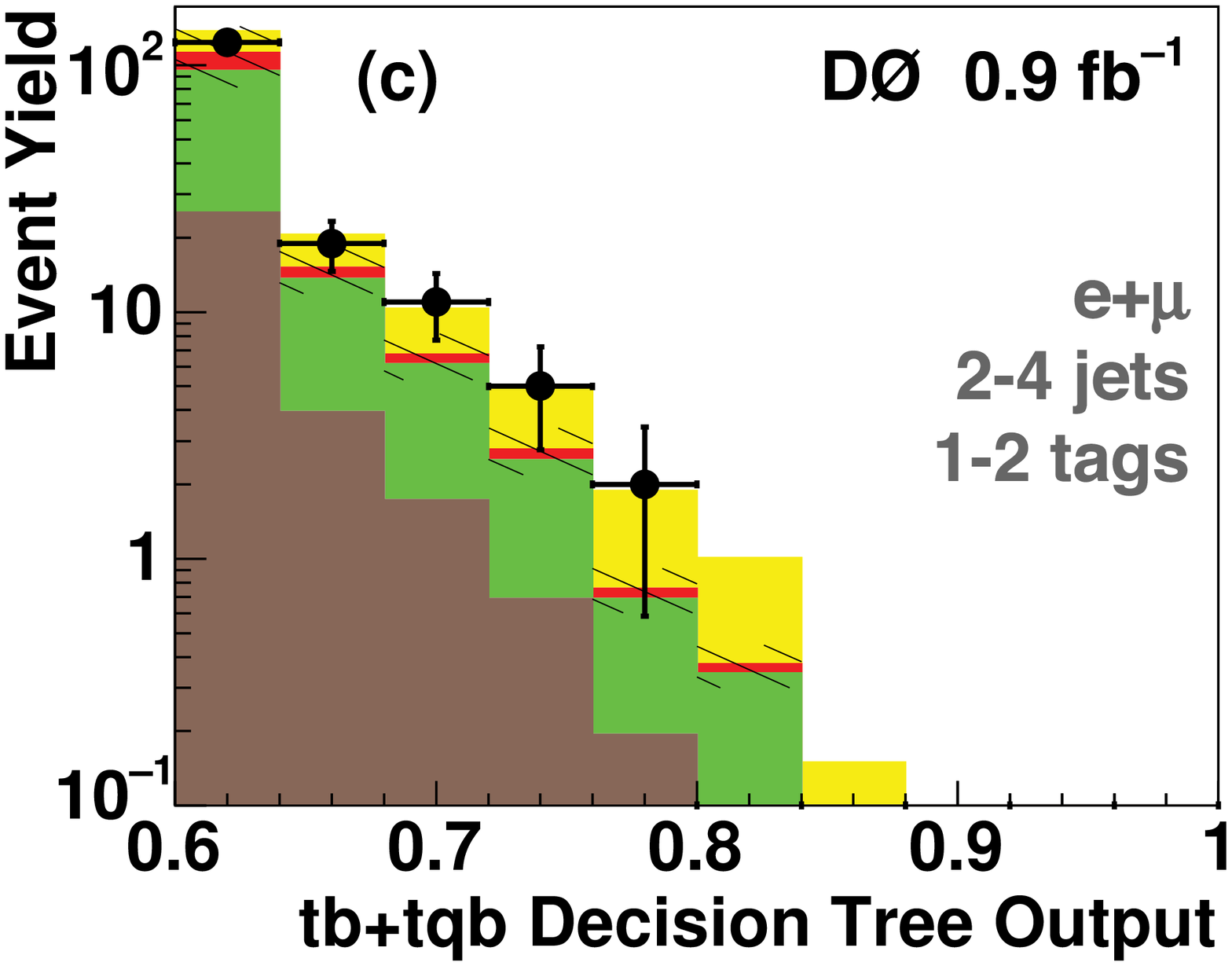}
\includegraphics[width=1.6in]{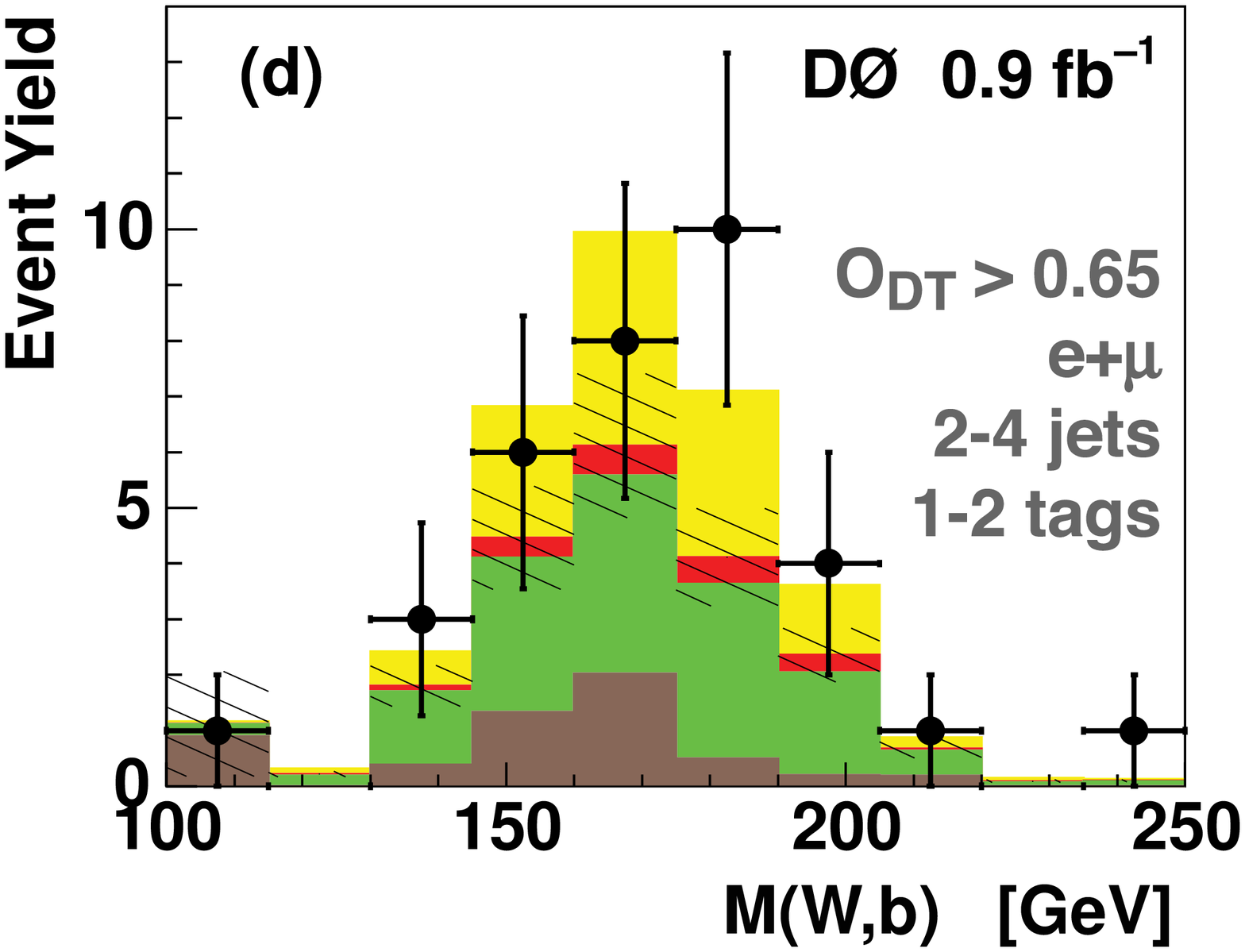}
\vspace{-0.1in}
\caption[dtplots]{Boosted decision tree output distributions for (a) a
$W$+jets-dominated control sample, (b) a {\ttbar}-dominated control
sample, and (c) the high-discriminant region of the sum of all 12
$tb$+$tqb$ DTs. For (a) and (b), $H_T = E_T^{\ell} + {\met}
+ \sum E_T^{\rm alljets}$. Plot (d) shows the invariant mass of the
reconstructed $W$~boson and highest-$p_T$ $b$-tagged jet for events
with $O_{\rm DT} > 0.65$. The hatched bands show the $\pm 1$ standard
deviation uncertainty on the background. The expected signal is shown
using the measured cross section.}
\label{dt-plots}
\vspace{-0.2in}
\end{figure}

We use a boosted decision tree (DT) in each of the twelve analysis
channels for three searches: $tb$+$tqb$, $tqb$, and $tb$. These 36 DTs
are trained to separate one of the signals from the sum of the
{\ttbar} and $W$+jets backgrounds. One-third of the MC signal and
background events is used for training; the remaining two-thirds are
used to determine the acceptances in an unbiased manner. A boosted
decision tree produces a quasi-continuous output distribution $O_{\rm
DT}$ ranging from zero to one, with background peaking closer to zero
and signal peaking closer to one. Figures~\ref{dt-plots}(a) and
\ref{dt-plots}(b) show the DT output distributions for two
background-dominated data samples to demonstrate the agreement between
background model and data. Figure~\ref{dt-plots}(c) shows the high end
of the sum of the 12 $tb$+$tqb$ DT outputs to illustrate where the
signal is expected, and Fig.~\ref{dt-plots}(d) shows the invariant
mass of the reconstructed $W$~boson with the highest-$p_T$ $b$-tagged
jet (where the neutrino longitudinal momentum has been chosen to be
the smaller absolute value of the two possible solutions to the mass
equation), for events in a signal-enhanced region with $O_{\rm DT} >
0.65$. The background peaks near the top quark mass because the DTs
select events similar to single top quark events.

We apply a Bayesian approach~\cite{bayes-limits} to measure the single
top quark production cross section. We form a binned likelihood as a
product over all bins and channels (lepton flavor, jet multiplicity,
and tag multiplicity) of the decision tree discriminant, separately
for the $tb$+$tqb$, $tqb$, and $tb$ analyses. We assume a Poisson
distribution for the observed counts and flat nonnegative prior
probabilities for the signal cross sections. Systematic uncertainties
and their correlations are taken into account by integrating over the
signal acceptances, background yields, and integrated luminosity with
Gaussian priors for each systematic uncertainty. The final posterior
probability density is computed as a function of the production cross
section. For each analysis, we measure the cross section using the
position of the posterior density peak and we take the $68\%$
asymmetric interval about the peak as the uncertainty on the
measurement.

We test the validity of the cross section measurement procedure using
six ensembles of pseudo-datasets selected from the full set of
$tb$+$tqb$ signal and background events weighted to represent their
expected proportions. A Poisson distribution with a mean equal to the
total number of selected events is randomly sampled to determine the
number of events in each pseudo-dataset. Each ensemble has a different
assumed $tb$+$tqb$ cross section between 2~pb and 8~pb. No significant
bias is seen in the mean of the measured cross sections for these
ensembles.

\begin{figure}[!h!tbp]
\includegraphics[width=2.2in]{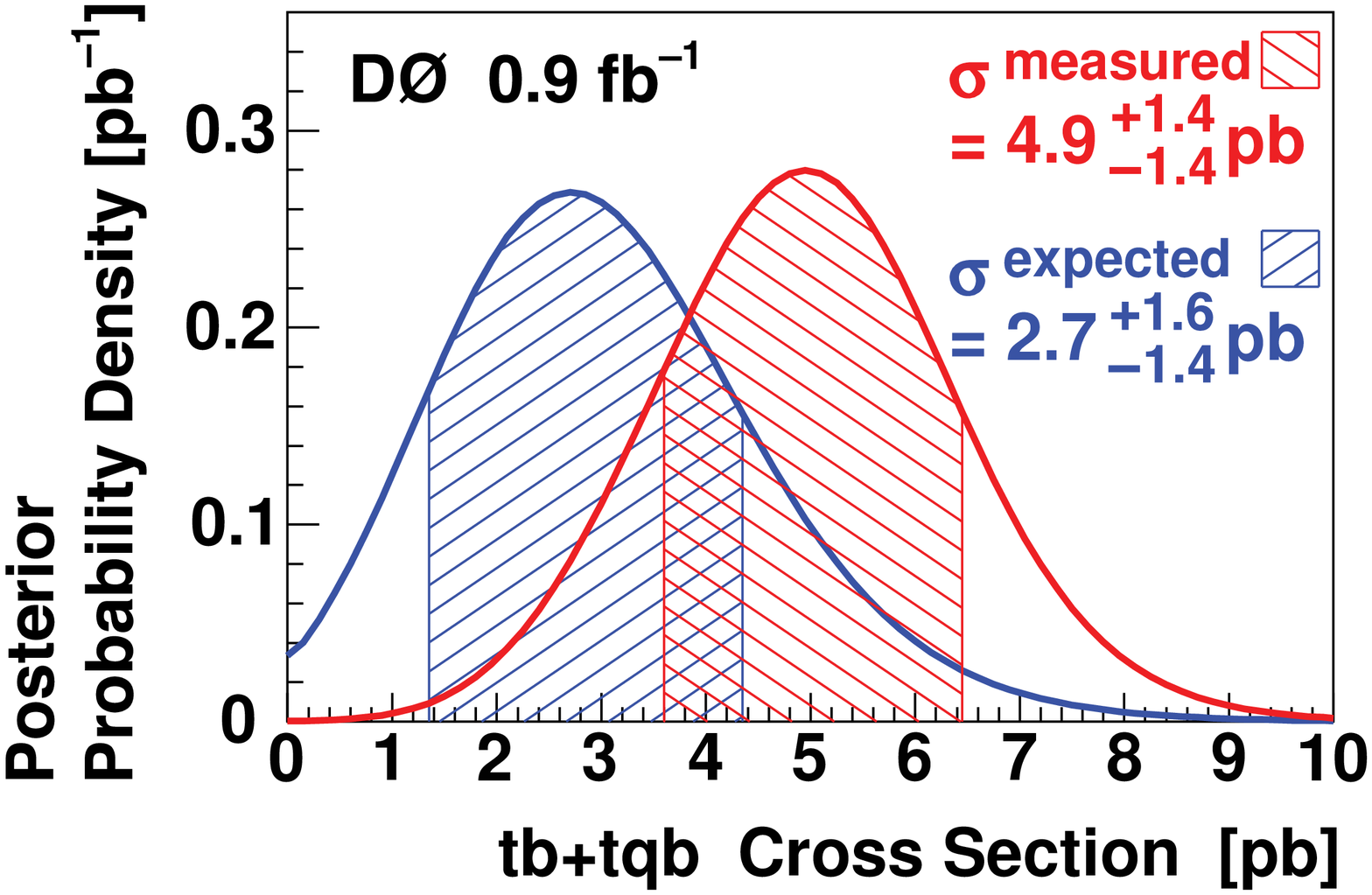}
\vspace{-0.1in}
\caption[1dposteriors]{Expected SM and measured Bayesian posterior
probability densities for the $tb$+$tqb$ cross section. The shaded
regions indicate one standard deviation above and below the peak
positions.}
\label{1d-posteriors}
\vspace{-0.1in}
\end{figure}

The expected SM and measured posterior probability densities for
$tb$+$tqb$ are shown in Fig.~\ref{1d-posteriors}. We use the measured
posterior density distribution for $tb$+$tqb$ as shown in
Fig.~\ref{1d-posteriors} and similar distributions for $tqb$ and $tb$
to make the following measurements:
$\sigma({\ppbar}{\rargap}tb+X,~tqb+X) = 4.9 \pm 1.4$~pb,
$\sigma({\ppbar}{\rargap}tqb+X) = 4.2 ^{+1.8}_{-1.4}$~pb, and
$\sigma({\ppbar}{\rargap}tb+X) = 1.0 \pm 0.9$~pb. These results are
consistent with the SM expectations. The uncertainties include
statistical and systematic components combined. The data statistics
contribute 1.2~pb to the total 1.4~pb uncertainty on the $tb$+$tqb$
cross section.

We assess how strongly this analysis rules out (or is expected to rule
out) the background-only hypothesis by measuring the probability for
the background to fluctuate up to give the measured (or SM) value of
the $tb$+$tqb$ cross section or greater. From an ensemble of over
68,000 background-only pseudo-datasets, with all systematic
uncertainties included, we find that the background fluctuates up to
give the SM cross section of 2.9~pb or greater $1.9\%$ of the time,
corresponding to an expected significance of 2.1 standard deviations
(SD) for a Gaussian distribution. The probability that the background
fluctuates up to produce the measured cross section of 4.9~pb or
greater is $0.035\%$, corresponding to a significance for our result
of 3.4~SD. Using a second ensemble of pseudo-datasets which includes a
SM $tb$+$tqb$ signal with 2.9~pb cross section, with all systematic
uncertainties included, we find the probability to measure a cross
section of at least 4.9~pb to be $11\%$.

We apply two alternative methods to calculate $tb$+$tqb$
discriminants. The first technique calculates the probability for each
event to be signal or background based on the leading-order matrix
element description of each process for two-jet and three-jet
events~\cite{matrix-elements}. It takes as input the four-momenta of
the reconstructed objects and incorporates the $b$-tagging information
for each event. This is a powerful method to extract the small signal
because it encodes the kinematic information of the signal and
background processes at the parton level. The probability that the
background fluctuates up to give the SM cross section or greater in
the matrix element analysis is $3.7\%$ (1.8~SD). We measure
$\sigma({\ppbar}{\rargap}tb+X,~tqb+X) = 4.6 ^{+1.8}_{-1.5}$~pb. The
probability for the background to fluctuate up to give a cross section
of at least 4.6~pb is $0.21\%$ (2.9~SD). The second alternative method
uses Bayesian neural networks~\cite{bayesianNNs} to separate
$tb$+$tqb$ signal from background. We train the networks separately
for each analysis channel on a sample of signal events and on an
equal-sized sample of background events containing the background
components in their expected proportions, using 24 input variables (a
subset of the 49 used in the boosted decision tree analysis). Large
numbers of networks are averaged, resulting in better separation than
can be achieved with a single network. The probability that the
background fluctuates up to give the SM cross section or greater in
the Bayesian neural network analysis is $9.7\%$ (1.3~SD). We measure
$\sigma({\ppbar}{\rargap}tb+X,~tqb+X) = 5.0 \pm 1.9$~pb. The
probability for the background to fluctuate up to give a cross section
of at least 5.0~pb is $0.89\%$ (2.4~SD).

The three analyses are correlated since they use the same signal and
background models and data, with almost the same systematic
uncertainties. We take the decision tree measurement as our main
result because this method has the lowest a priori probability for the
background to have fluctuated up to give the SM cross section or
greater. That is, we expect the decision tree analysis to rule out the
background-only hypothesis with greatest significance.

We use the decision tree measurement of the $tb$+$tqb$ cross section
to derive a first direct measurement of the strength of the $V$$-$$A$
coupling $|V_{tb} f_1^L|$ in the $Wtb$ vertex, where $f_1^L$ is an
arbitrary left-handed form factor~\cite{vtb-theory}. We measure
$|V_{tb} f_1^L| = 1.3 \pm 0.2$. This measurement assumes $|V_{td}|^2 +
|V_{ts}|^2 \ll |V_{tb}|^2$ and a pure $V$$-$$A$ and CP-conserving
$Wtb$ interaction. Assuming in addition that $f_1^L = 1$ and using a
flat prior for $|V_{tb}|^2$ from 0 to 1, we obtain $0.68 < |V_{tb}|
\le 1$ at $95\%$~C.L. These measurements make no assumptions about the
number of quark families or CKM matrix unitarity.

To summarize, we have performed a search for single top quark
production using 0.9~fb$^{-1}$ of data collected by the D0 experiment
at the Tevatron collider. We find an excess of events over the
background prediction in the high discriminant output region and
interpret it as evidence for single top quark production. The excess
has a significance of 3.4 standard deviations. We use the boosted
decision tree discriminant output distributions to make the first
measurement of the single top quark cross section:
$\sigma({\ppbar}{\rargap}tb+X,~tqb+X) = 4.9 \pm 1.4$~pb. We use this
cross section measurement to make the first direct measurement of the
CKM matrix element $|V_{tb}|$ without assuming CKM matrix unitarity,
and find $0.68 < |V_{tb}| \le 1$ at $95\%$~C.L.

%---------------------------------------------------------------------

We thank the staffs at Fermilab and collaborating institutions, 
and acknowledge support from the 
DOE and NSF (USA);
CEA and CNRS/IN2P3 (France);
FASI, Rosatom and RFBR (Russia);
CAPES, CNPq, FAPERJ, FAPESP and FUNDUNESP (Brazil);
DAE and DST (India);
Colciencias (Colombia);
CONACyT (Mexico);
KRF and KOSEF (Korea);
CONICET and UBACyT (Argentina);
FOM (The Netherlands);
PPARC (United Kingdom);
MSMT (Czech Republic);
CRC Program, CFI, NSERC and WestGrid Project (Canada);
BMBF and DFG (Germany);
SFI (Ireland);
The Swedish Research Council (Sweden);
The Radcliffe Institute Fellowship program;
Research Corporation;
Alexander von Humboldt Foundation;
and the Marie Curie Program.

\vspace{-0.25in}

%---------------------------------------------------------------------
%---------------------------------------------------------------------

%---------------------------------------------------------------------
%---------------------------------------------------------------------

\end{document}